# Serotonin as a Creativity Pump

Tariq Khan*

June 16, 2024

### Abstract

The location of Western Civilization defined nations in Europe and in the United States within the largest global pollen environments on the planet is proposed as a key factor leading to their success. Environments with dense pollen concentrations will cause large up and down changes in serum histamine directly causing reductions and increases in brain serotonin i.e., a larger "serotonin slope," linked to higher levels of creativity. The pollen ecosystem in northern latitude nations is thus considered the hidden driver of the success of these populations as the biochemical interaction between histamine and serotonin leads to a "creativity pump" that is proposed as the fundamental driver of intelligence in micro and macro human populations.

___________________________________________________________________

University of Nebraska at Omaha. Email. tariqkhan@unomaha.edu ORCID: 0000-0002-3236-0786

# 1 Introduction

> *"Other, less abstract approaches to improving creativity center around the importance of serotonin. According to research… serotonin levels are tied to creativity… a gene pertaining to serotonin, known as TPH1, is associated with "figural" creativity - or creativity regarding shapes, diagrams, and drawings."*
> ______________________________________________
> Jandy Le and Michael Xiong, *The Scientific Origin of Creativity*

Historians have attempted to identify the causes for the success of "Western" nations over other regions of the world especially as it relates to the accumulation of technology and knowledge and their eventual power in the modern world. False factors like race or religion and physical factors like geography, weather, and even luck (which do have their place) have all been part of the analysis. But, perhaps, there is another factor that observational data appears to support. Taken as a given that increased changes in brain serotonin correlate with increased creativity, then regions of the globe with increased pollen density will lead to increased up and down changes in serum histamine that will thus also lead to increased changes in brain serotonin (a larger "serotonin slope") and thus, possibly, create an unseen, if not unrealized, benefit of an environment situated for improved creativity. We thus have perhaps an additional driver behind the success of specific nations of the world and even aggregate modern human history.

One could even speculate that the very act of domestication into living areas e.g., caves, dwellings, etc. began the initial "serotonin slope" leading to the origination of human intelligence. The increased evidence for intelligence in domesticated dogs would further reinforce this proposal.

# 2 Discussion

An argument can be made that any correlation between pollen level and societies is a result of coincidence as societies could, over time, tend to settle in areas where agriculture would flourish e.g., warm temperate forests or temperate forests (Figure 1), however human beings did not originate from these areas, and we evolved and migrated from the Africa tropics.

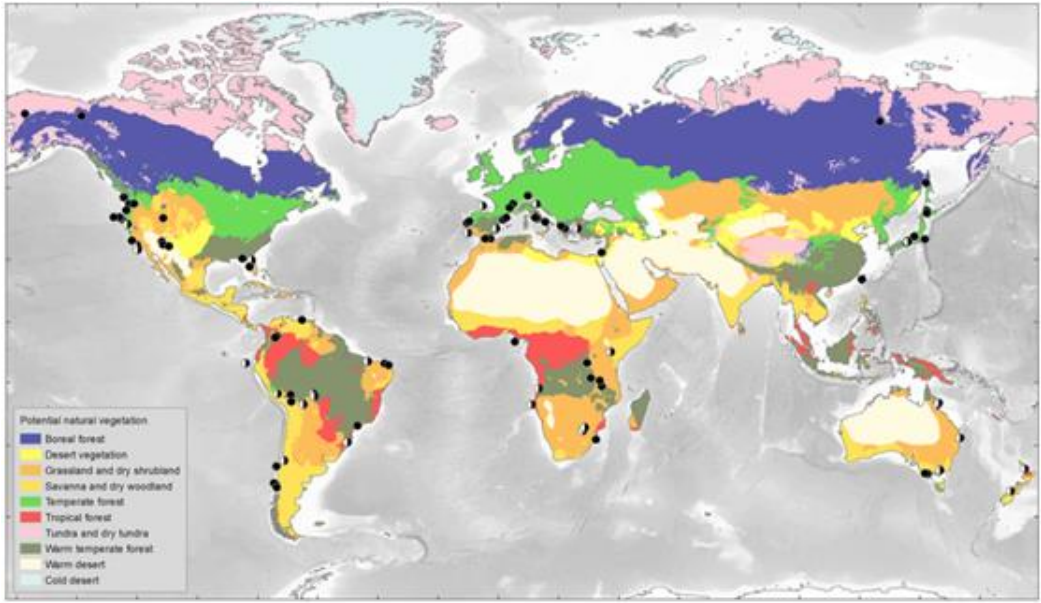

Figure 1. Map of the locations of pollen sites from the last glacial ice age period (Goni, 2017).

Recent research, including maps of present day and historical pollen density, matches the proposed theory. Manuel Chevalier et al., in their 2020 paper Pollen-based climate reconstruction techniques for late Quaternary studies from Earth-Science Reviews includes a map of different pollen samples (Figure 2) that shows a frequency distribution whose density is nearly identical to that of the map of powerful modern nations including the United States, Canada, Europe, China, and Japan. Their work also includes a map of historic pollen records again with densities or locations correlating to the proposed model (Figure 3). Note the mapped correlation to modern nation-states with the greatest technology, industry, and success.

Kuan-Wei Chen et al in the 2018 edition of the International Archives of Allergy and Immunology included a global map of ragweed in their paper Ragweed Pollen Allergy: Burden, Characteristics, and Management of an Imported Allergen Source in Europe. Again, we see the location and density of an example pollen, in this case fall ragweed, correlating to nations that are major modern and technological world powers (Figure 4).

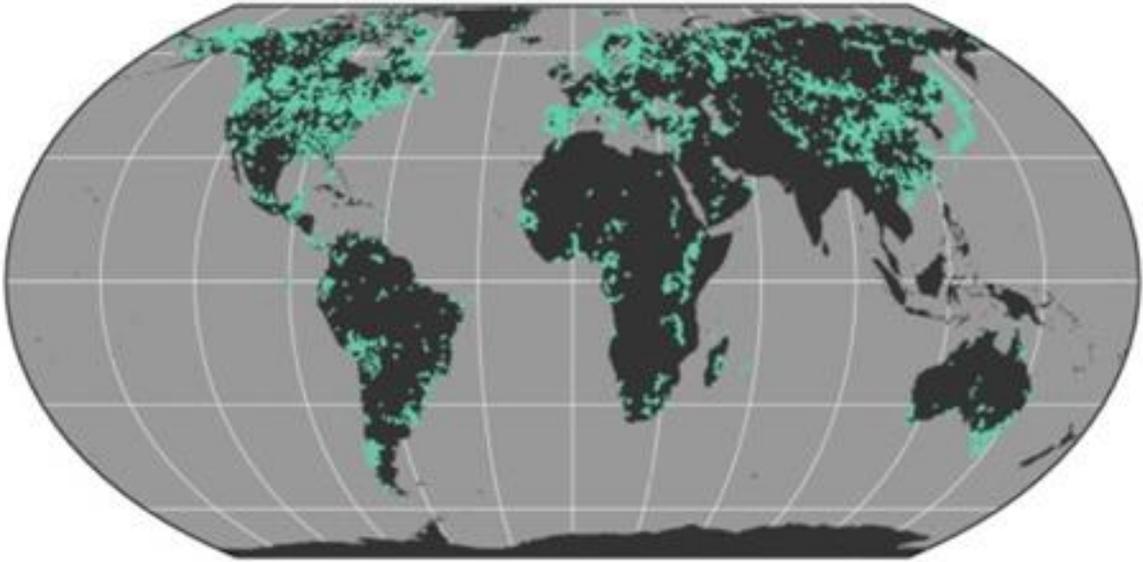

Figure 2. Map showing locations of significant pollen samples (green) from modern testing (Chevalier, 2020).

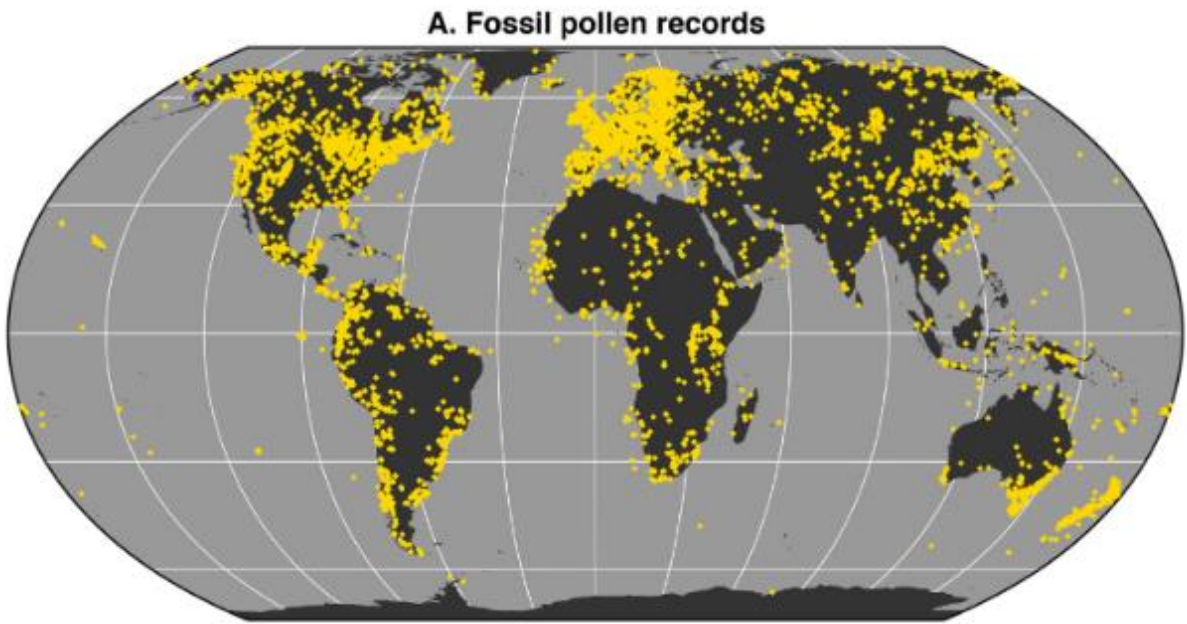

Figure 3. Map of ancient pollen locations (yellow) from fossil records (Chevalier, 2020).

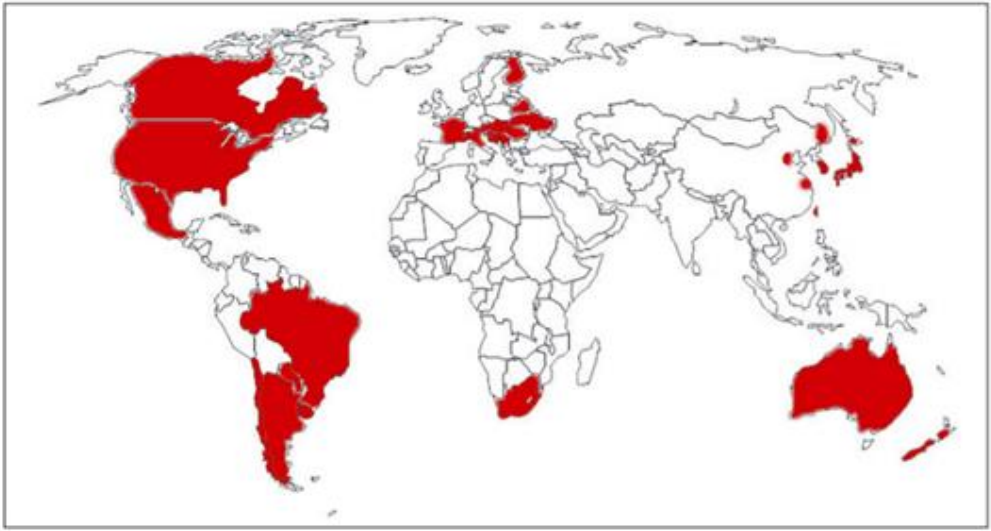

Figure 4. Map showing modern-day worldwide distribution of ragweed pollen by country (red) (Chen, 2018).

Jonathan Storkey et al., in the 2014 edition of PLoS ONE paper called A Process-Based Approach to Predicting the Effect of Climate Change on the Distribution of an Invasive Allergenic Plant in Europe, show a map of even greater detail in central Europe noting density of ragweed (Figure 5). This same map overlaid on a modern political map of Europe (Figure 6) shows a significant association with pollen density and the major cities known as the "intellectual capitals" of Europe. We see the pollen density associated with intellectual European capitals including Paris (Pablo Picasso, Vincent van Gogh, Marie Curie, etc. Figure 7), Vienna (Kurt Gödel, Ernst Mach, Ludwig Boltzmann, Sigmund Freud, etc. Figure 8), Budapest (John Von Neumann), the German cities Munich (Albert Einstein), Salzburg (Wolfgang Amadeus Mozart), and Nuremburg (Gottfried Wilhelm Leibniz) (Figure 9), Milan (Leonardo Da Vinci, etc. Figure 10), Zurich (Carl Jung), etc. Cities, especially on the border of the dense ragweed pollen regions, would have the largest "serotonin slopes" and thus produce individual and aggregate populations, as noted, with higher levels of creativity and intellectual performance (Figure 11).

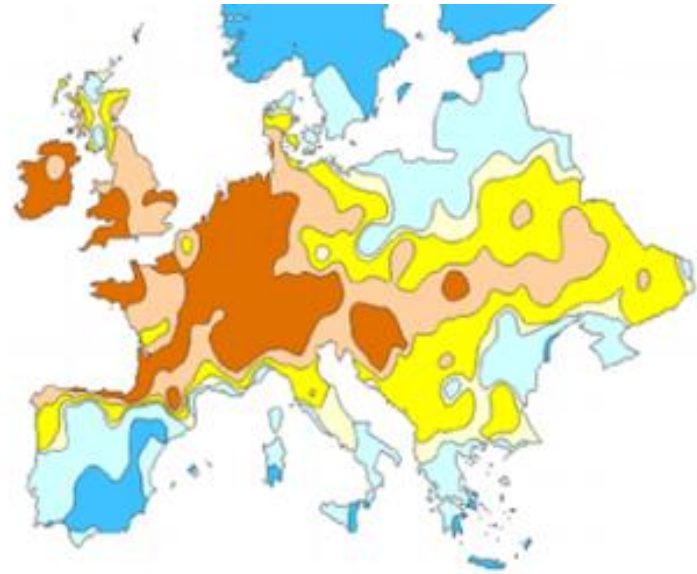

Figure 5. Map showing density of ragweed pollen over continental Europe (Storkey, 2014).

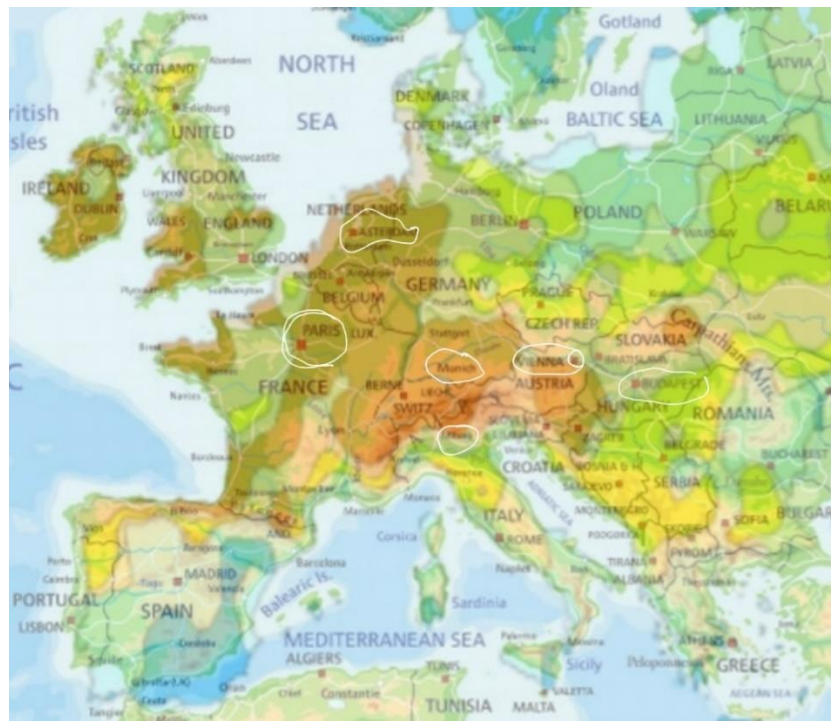

Figure 6. Map of ragweed distribution in Europe overlaid on political map. The key "intellectual capitals" of Europe lie within the dense ragweed pollen zones including Paris, Amsterdam, Munich, Vienna, Milan, and Budapest (Storkey, 2014).

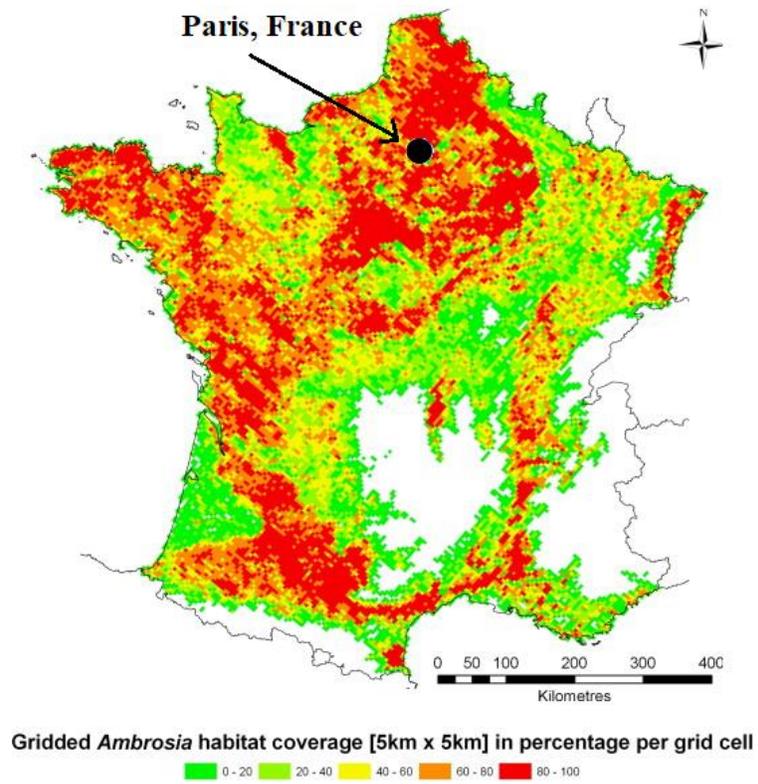

Figure 7. Modern map of ragweed pollen density in France (Thibaudon, 2014).

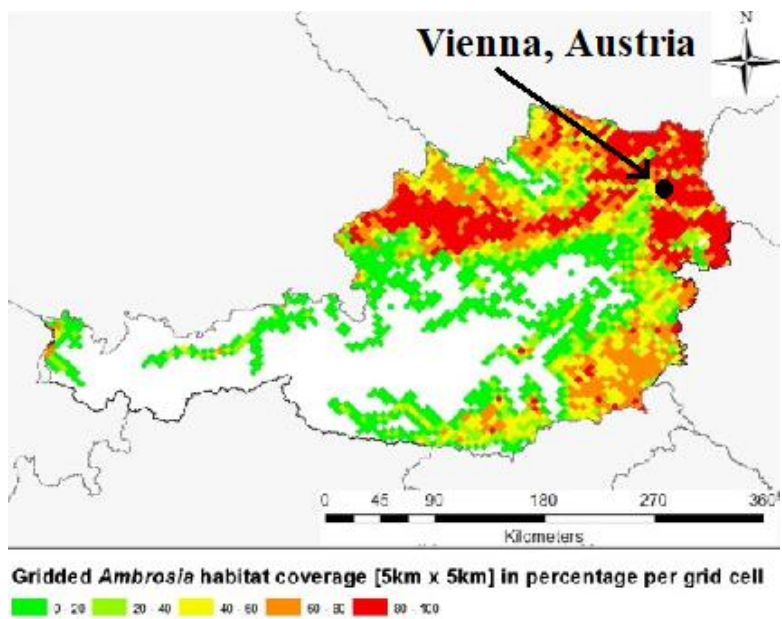

Figure 8. Modern map of Austria showing ragweed pollen density (Karrer, 2015).

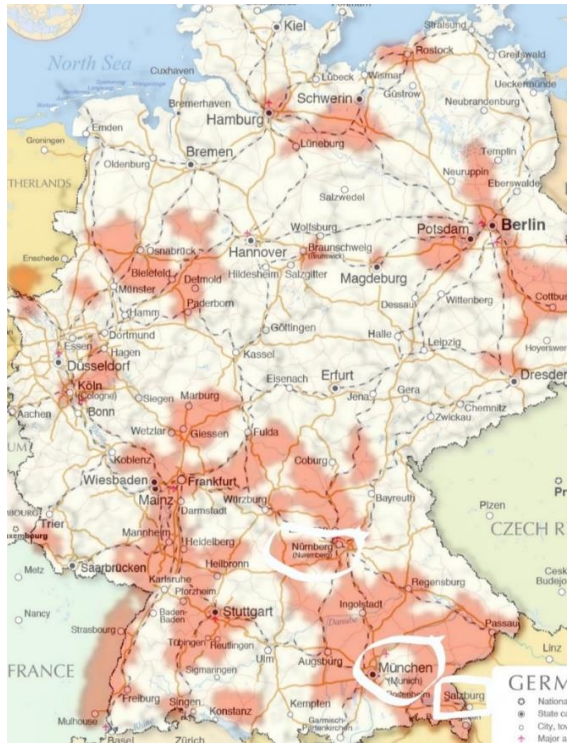

Figure 9. Map showing pollen dense regions of Germany with "intellectual capitals" of Nuremberg, Munich, and Salzburg located in the densest regions (Buters, 2015).

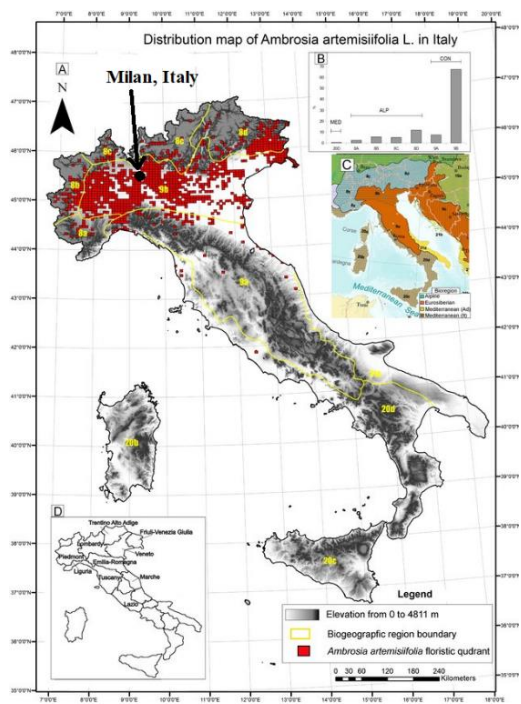

Figure 10. Modern map of ragweed pollen density of Italy with Milan in the center of the zone (Bonini, 2017).

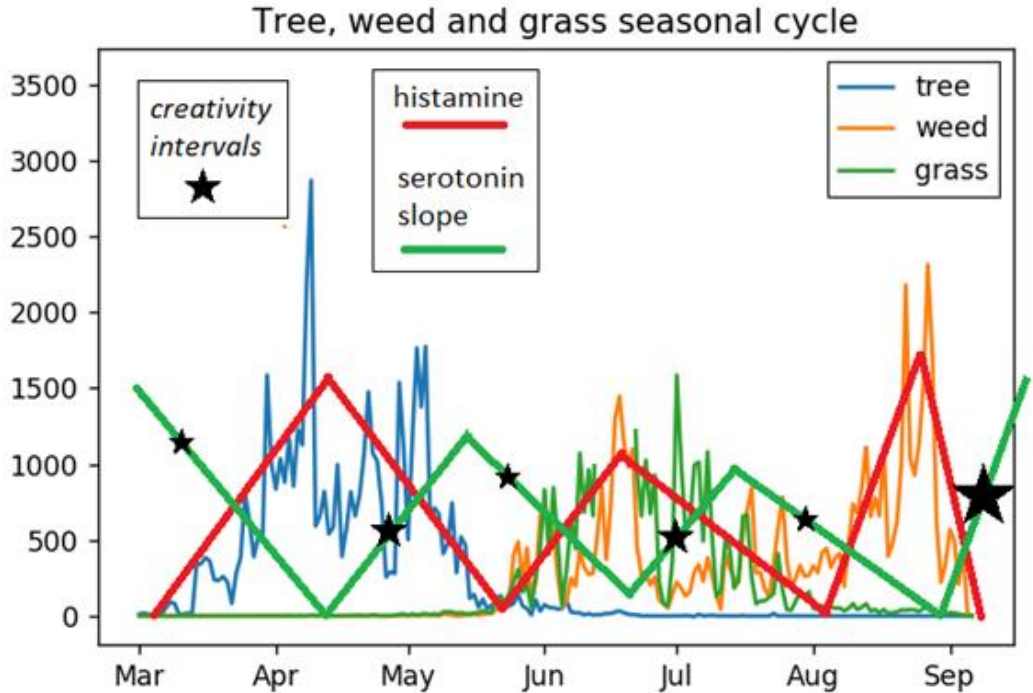

Figure 11. Graph of the human "serotonin slope" driving creativity versus spring tree, summer grass, and fall ragweed pollen spikes (Cowie, 2018).

## 3   The Genius of Newton

Of note, is the absence of Isaac Newton from the genius producing cities. How do we explain one of the greatest geniuses of all time with London not in the area of the model? But Isaac Newton does fit in the model, one must simply know the personal history of Newton. There is zero doubt about the scale and quality of his genius but of interest for this model is "where" did he do the majority of great and original work? When the Black Death plague hit London, Newton isolated himself for years in the countryside at his Woolsthorpe Manor. It was here (amongst the pollen) that he made his major discoveries - unknowingly placing himself into a high "serotonin slope" environment versus especially mugwort pollen. The British weed in the same ragweed family is called mugwort. Note the overlay map of the mugwort density of London versus his countryside manor in Woolsthorpe (Figure 12) and note the literal weeds, trees, and grass there even still today (Figure 13).

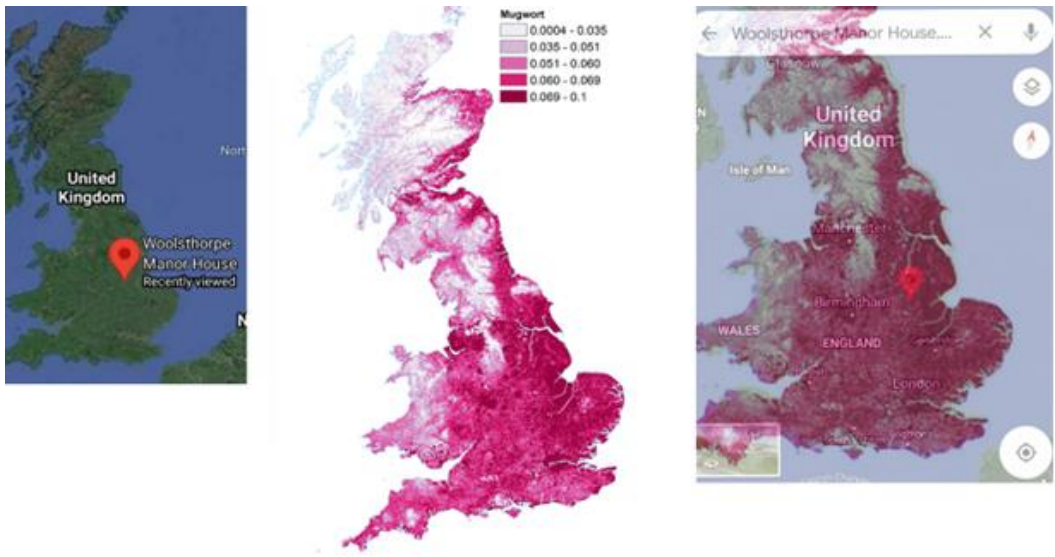

Figure 12. Maps of the United Kingdom and ragweed pollen density noting the location of Isaac Newton's summer manor house in Woolsthorpe deep in the dense region (Forster, 2019).

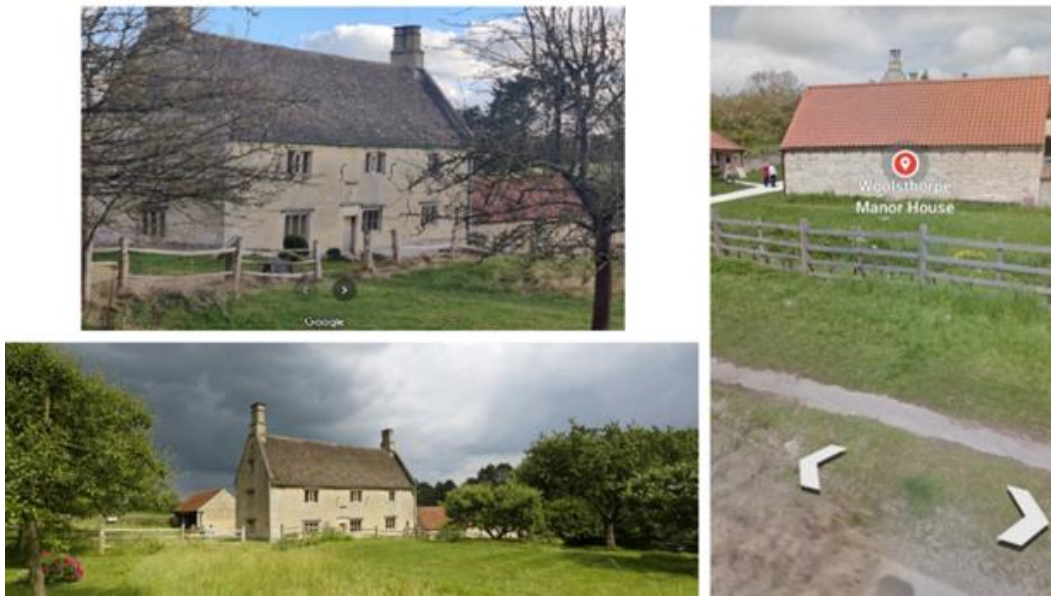

Figure 13. The countryside home of Isaac Newton for two years during the plague was in Woolsthorpe Manor, a region surrounded by dense weed and tree pollen, where his creativity shined, and Newton developed his Calculus, optics, and law of Gravitation (Wikipedia, 2024).

As noted in Wikipedia:
> "Soon after Newton had obtained his BA degree in August 1665, the university temporarily closed as a precaution against the Great Plague. Although he had been undistinguished as a Cambridge student, Newton′s private studies at his home in Woolsthorpe over the subsequent two years saw the development of his theories on calculus, optics, and the law of gravitation."

# 4 Conclusion

One may argue that other ancient river-based civilizations like ancient Egypt, Inca, Mexico, Babylon, Angor Wat, etc. also had complex societies with intellectual achievements, however those civilizations took hundreds and thousands of years to mature and develop while the increase in intellectual creativity and innovative and thus technology and power of modern "Western Civilization" nations and China and Japan (Figure 14) moved at an exponentially higher rate of growth by comparison. This paper attempts to explain the cause of that rate of intellectual and technological advancement via the environmental and human bio-chemical mechanism and thus unbeknownst location driven accident. In summary, the biochemical interaction between histamine and serotonin leads to a ″creativity pump″ that is the fundamental driver of intelligence in micro and macro human populations.

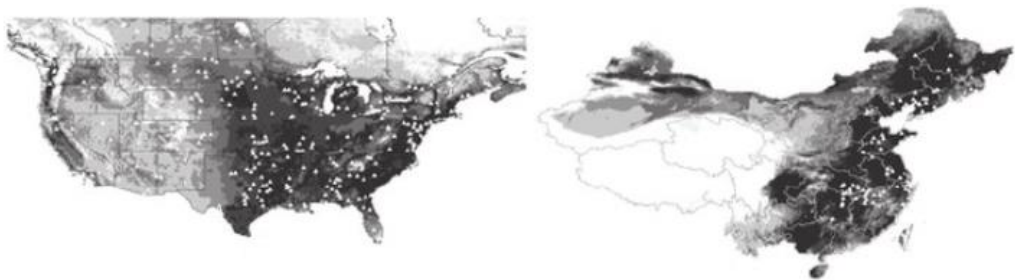

Figure 14. Modern-day density maps of ragweed pollen in the United States and China (Chen, 2007).

From the perspective of an applied approach to this theory, a mild antihistamine taken any time of the year like chlorphenamine (a mild SSRI with an over-the-counter cost of about a tenth of a penny) if taken every other day, or perhaps for a few days on and then a few days off, to induce or stimulate the up-and-down "serotonin slope" (compare as an analogy versus Tesla's back and forth alternating current (AC) used for electric power generation), may be just enough varying change in brain serum serotonin to simulate a similar histamine/serotonin "creativity pump" in any person's brain similar to that observed in nature and described in this paper. One might even speculate as to the benefit of humans living or working in high-rise or top floor skyscrapers (or castles) as beneficial to creativity as the literal change from a low environment, surrounded by pollen, to the high region, away from allergens, could also create a similar, if not mild, "creativity pump." A similar scenario might be on-and-off exercise intervals that boost serotonin.

One might speculate further noting the interesting correlation between the location of the first works of human art or drawings and jewelry in the Blombos Cave in South Africa and its location having very high density of pollen per fossil record data (Figures 15-17). Per the described model, it does not matter if ragweed and humans both "prefer" areas that are low, moist, and near bodies of water. The point is that it was as soon as early humans migrated and settled in those pollen (thus histamine and serotonin) dense regions, that the "brain spark" from the "serotonin slope," leading to creativity, begins. Note that fossil records show dense pollen in Africa primarily in two areas: in South Africa and Tanzania (Figure 17). Thus, it should come as no surprise that in the Blombos Cave near Cape Town in South Africa, we see the very first cave drawings and jewelry i.e., brains using analogy and creativity (Figure 15). To summarize, humans originate in Botswana and migrate to the low wet water plentiful areas (Figure 16). They domesticate (in the sense of sheltering from their environment) and they also change environments to one where their immune systems are not familiar (traveling adults), and they encounter unfamiliar weeds with pollen. This caused the biochemical histamine reactions and, thus, the discussed up-and-down changes in brain serum serotonin and thus human creativity is born.

In conclusion, if this described model does represent the mechanism for creativity in brains (in essence intelligence) then further researchers can "fine tune" the details. They can identify exactly how much serotonin change is needed, and over how much time, for "sparks of creativity" to occur and exactly how the deeper process works. Then they cannot only optimize the technique - envision one million students taking 1 mg of chlorphenamine every other day boosting grades - they may also now finally understand (monitor via PET scans and real-time functional MRIs) how the complete brain and broader intelligence fully works. In essence, this model can be a spotlight or "step in the right direction" to then lead to the actual reverse engineering of the human brain, and all its functional areas via these histamine and serotonin interactions, and thus gain additional or critical progress toward the creation of true general artificial intelligence machines.

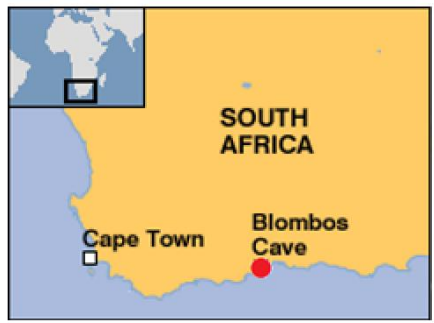

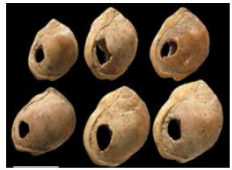

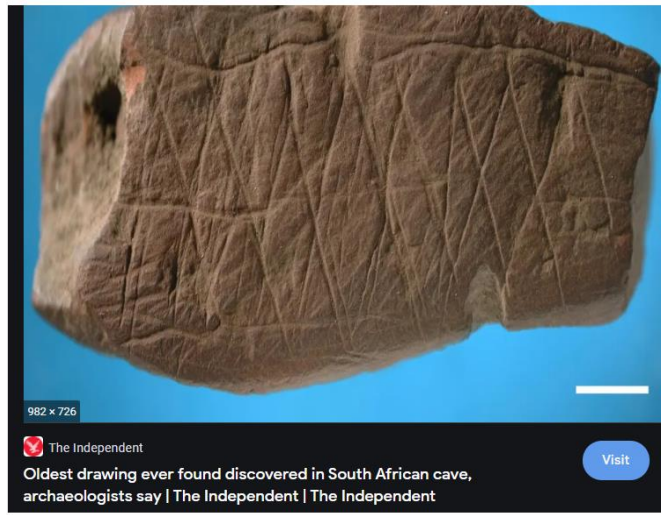

Figure 15. Map of the location of the first drawings and jewelry from ancient humans in South Africa (Gabbatiss, 2018).

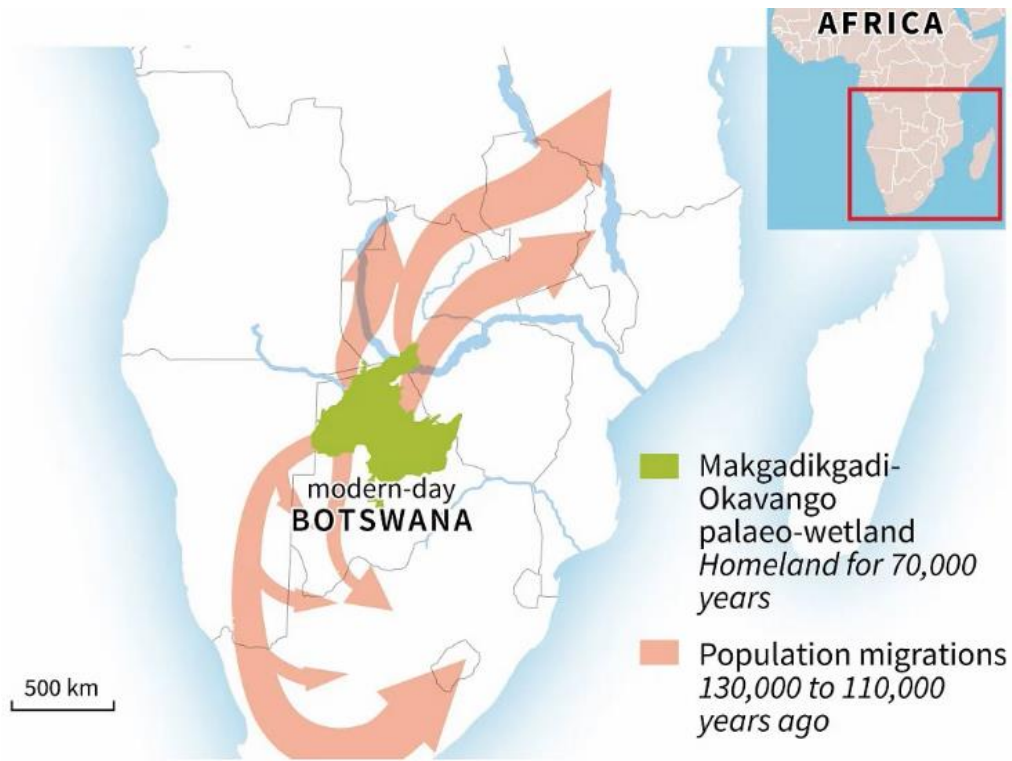

Figure 16. Ancient human migrations from our primordial origins in central Africa (Staff, 2019).

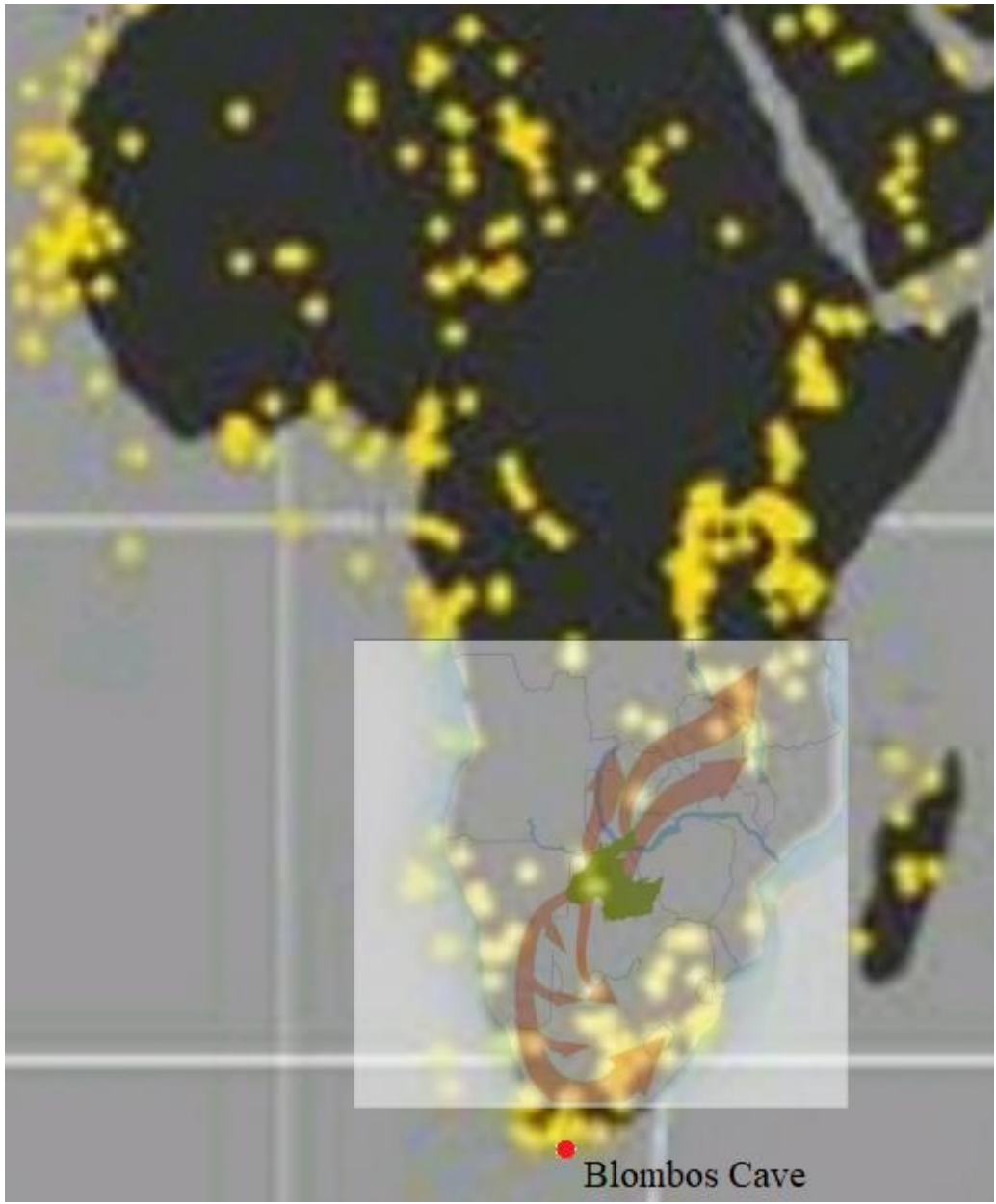

Figure 17. Overlay of human migration map and fossil record locations of pollen in Africa (Staff 2019).

# References

Allergy cells in the rodent brain may keep baseline anxiety under control. (2008). Immune to Anxiety. *Science*. https://www.science.org/content/article/immune-anxiety

Amos, Jonathan. (2004). Cave yields 'earliest jewelry'. BBC News Online. http://news.bbc.co.uk/2/hi/science/nature/3629559.stm

Bonini, M. et al. (2017). Ambrosia pollen source inventory for Italy: a multi-purpose tool to assess the impact of the ragweed leaf beetle (Ophraella communa LeSage) on populations of its host plant. *International Journal of Biometeorology,* 62, 597-608.

Chen, Hao, Chen, Lijun & Albright, Thomas. (2007). Developing Habitat-suitability Maps of Invasive Ragweed (Ambrosia artemisiifolia.L) in China Using GIS and Statistical Methods. 10.1007/978-3-540-71318-0_8.

Buters, J., et al. (2015) *Ambrosia artemisiifolia* (ragweed) in Germany - current presence, allergological relevance and containment procedures. Allergo J Int. 2015; 24: 108-120. doi: 10.1007/s40629-015-0060-6. Epub 2015 Jul 11. PMID: 27226949; PMCID: PMC4861741.

Chen, Kuan-Wei et al. (2018). Ragweed Pollen Allergy: Burden, Characteristics, and Management of an Imported Allergen Source in Europe. *International Archives of Allergy and Immunology*. https://www.karger.com/Article/Pdf/487997

Chevalier, Manuel et al. (2020). Pollen-based climate reconstruction techniques for late Quaternary studies. *Earth-Science Reviews*, Volume 210, 103384, ISSN 0012-8252. https://www.sciencedirect.com/science/article/pii/S001282522030430X

Cowie, Sophie, Arthur, Rudy & Williams, Hywel. (2018). Tracking Pollen and Hayfever in the UK Using Social Media. *Sensors*. 18. 4434. 10.3390/s18124434. https://www.researchgate.net/publication/329669438_choo_Tracking_Pollen_and_Hayfever_in_the_UK_Using_Social_Media/figures?lo=1

Forster, K. (2017, May 19). Hay fever maps: Which areas to avoid where the pollen concentrations are highest. *Independent*. https://www.independent.co.uk/news/health/hay-fever-maps-which-areas-avoid-pollen-allergy-concentrations-highest-a7744351.html

Gabbatiss, J. (2018). Oldest drawing ever found discovered in South African cave, archaeologists say. *Independent*. https://www.independent.co.uk/news/science/archaeology/oldest-drawing-ever-south-africa-blombos-cave-art-hashtag-rock-ochre-a8534696.html

Goni, Sanchez et al. (2017). The ACER pollen and charcoal database: A global resource to document vegetation and fire response to abrupt climate changes during the last glacial period. *Earth System Science Data*. 9. 679-695. 10.5194/essd-9-679-2017. https://www.researchgate.net/figure/Map-with-location-of-the-93-marine-and-terrestrial-pollen-sites-covering-part-of-or-all_fig2_319631062

Karrer, G. et al. (2015). Ragweed (Ambrosia) pollen source inventory for Austria. *Science of The Total Environment*, 523, 120-128. https://doi.org/10.1016/j.scitotenv.2015.03.108.

Le, Jandy, Xiong, Michael, & Joshi Jwalin. (2019). The Scientific Origin of Creativity. *Neurotech@Berkeley*. https://medium.com/neurotech-berkeley/the-scientific-origin-of-creativity-587799f0fbe2

Nautiyal, K. M., Dailey, C. A., Jahn, J. L., Rodriquez, E., Son, N. H., Sweedler, J. V., & Silver, R. (2012). Serotonin of mast cell origin contributes to hippocampal function. *The European Journal of Neuroscience*, 36(3), 2347–2359. https://doi.org/10.1111/j.1460-9568.2012.08138.x https://www.ncbi.nlm.nih.gov/pmc/articles/PMC3721752/

Ratner, P. (2021). From 1.8 million years ago, earliest evidence of human activity found. *BIGThink.com*. https://bigthink.com/the-past/earliest-evidence-human-activity/?fbclid=IwAR0Jf1L_R6nH_6w3wEqvJJqX3FUg0GKToLxK-BR_-9VzvJKKQSZLIbHT7i4

Shaar, R. et al. (2021). Magnetostratigraphy and cosmogenic dating of Wonderwerk Cave: New constraints for the chronology of the South African Earlier Stone Age. *Quaternary Science Reviews*, 259, 106907. https://doi.org/10.1016/j.quascirev.2021.106907. https://www.sciencedirect.com/science/article/pii/S0277379121001141

Staff Correspondent. (2019). Humanity's homeland found in ancient Botswana. *Prothom Alo*. https://en.prothomalo.com/science-technology/Humanity-s-homeland-found-in-ancient-Botswana

Storkey, Jonathan, et al. (2014). A Process-Based Approach to Predicting the Effect of Climate Change on the Distribution of an Invasive Allergenic Plant in Europe. *PLOS ONE,* 9(2): e88156. https://doi.org/10.1371/journal.pone.0088156

Theoharides, T., Bondy, P., & Tsakalos, N. et al. (1982). Differential release of serotonin and histamine from mast cells. *Nature,* 297, 229–231. https://doi.org/10.1038/297229a0 https://www.nature.com/articles/297229a0

Thibaudon, Michel et al. (2014). Ragweed pollen source inventory for France – The second largest centre of Ambrosia in Europe, Atmospheric Environment, Volume 83, Pages 62-71, ISSN 1352-2310, https://doi.org/10.1016/j.atmosenv.2013.10.057.

Wikipedia contributors. (2024, June 10). Isaac Newton. In Wikipedia, The Free Encyclopedia. Retrieved 22:55, June 16, 2024, from https://en.wikipedia.org/w/index.php?title=Isaac_Newton&oldid=1228344161

Wikipedia contributors. (2021). Olduvai Gorge. Wikipedia, The Free Encyclopedia. https://en.wikipedia.org/w/index.php?title=Olduvai_Gorge&oldid=1059721958